\documentclass[11pt]{article}
\usepackage{fullpage}

\usepackage{latexsym}
\usepackage{amssymb}
\usepackage{amsmath}
\usepackage{amsthm}
\usepackage{amsfonts}
\usepackage{dsfont}
\usepackage{mathtools}

\usepackage{booktabs}
\usepackage{enumitem}
\usepackage{graphicx}
\usepackage{xcolor}
\usepackage{hyperref}

\newtheorem{theorem}{Theorem}

\newtheorem{Proposition}[theorem]{Proposition}

\newtheorem{Definition}[theorem]{Definition}

\newtheorem{Remark}[theorem]{Remark}

\newtheorem{Conjecture}[theorem]{Conjecture}

\newcommand{\cC}{\ensuremath{\mathcal C}}
\newcommand{\cD}{\ensuremath{\mathcal D}}
\newcommand{\cE}{\ensuremath{\mathcal E}}

\newcommand{\cG}{\ensuremath{\mathcal G}}

\newcommand{\cN}{\ensuremath{\mathcal N}}

\newcommand{\cV}{\ensuremath{\mathcal V}}

\newcommand{\bbN}{{\ensuremath{\mathbb N}} }

\title{A Phase Transition for Opinion Dynamics with Competing Biases}
\author{Federico Capannoli \and 
Emilio Cruciani \and 
Hlafo Alfie Mimun \and 
Matteo Quattropani}
\date{\small 
Leiden University \hspace{1em} 
European University of Rome \hspace{1em} 
John Cabot University \hspace{1em} 
Roma Tre University}

\begin{document}

\maketitle

\begingroup
\renewcommand\thefootnote{}%
\footnotetext{\tiny Email addresses: \url{f.capannoli@math.leidenuniv.nl},
\url{emilio.cruciani@unier.it},
\url{hlafo.mimun@johncabot.edu},
\url{matteo.quattropani@uniroma3.it}}%
\endgroup

\begin{abstract}
We study a nonlinear dynamics of binary opinions in a population of agents connected by a directed network, influenced by two competing forces. 
On the one hand, agents are \emph{stubborn}, i.e., have a tendency for one of the two opinions; on the other hand, there is a \emph{disruptive bias}, $p\in[0,1]$, that drives the agents toward the other opinion.
The disruptive bias models external factors, such as market innovations or social controllers, aiming to challenge the status quo, while agents' stubbornness reinforces the initial opinion making it harder for the external bias to drive the process toward change.
Each agent updates its opinion according to a nonlinear function of the states of its neighbors and of the bias $p$.
We consider the case of random directed graphs with prescribed in- and out-degree sequences and we prove that the dynamics exhibits a phase transition: when the disruptive bias $p$ is larger than a critical threshold $p_c$, the population converges in constant time to a consensus on the disruptive opinion. 
Conversely, when the bias $p$ is less than $p_c$, the system enters a metastable state in which only a fraction of agents $q_\star(p)<1$ will share the new opinion for a long time. We characterize $p_c$ and $q_\star(p)$ explicitly, showing that they only depend on few simple statistics of the degree sequences.
Our analysis relies on a dual system of branching, coalescing, and dying particles, which we show exhibits equivalent behavior and allows a rigorous characterization of the system's dynamics.
Our results characterize the interplay between the degree of the agents, their stubbornness, and the external bias, shedding light on the tipping points of opinion dynamics in networks.
\end{abstract}

\section{Introduction}\label{sec:intro}

Modeling the evolution of opinions in social and technological systems is a fundamental problem at the intersection of artificial intelligence and network science \cite{castellano2009statistical,newman2018networks}. 
In particular, understanding how collective behaviors emerge from individual interactions is key to predicting and influencing processes such as the adoption of new technologies \cite{montanari2010spread} or the spread of misinformation \cite{del2016spreading}. 
Herein, we propose and analyze a nonlinear opinion dynamics model that captures the competition between two opposing forces: an external \textit{disruptive bias}, which pushes the system toward a new opinion, and individuals' \textit{stubbornness}, which weakens the effect of the external disruptive bias by reinforcing the old opinion.

The two competing biases give rise to a complex interplay between endogenous and exogenous forces in the system, which is ubiquitous in real-world systems. 
The disruptive bias can represent external interventions such as targeted marketing campaigns, technological innovations, or coordinated disinformation efforts \cite{montanari2010spread,anagnostopoulos2022biased,lesfari2022biased}. 
Stubbornness, on the other hand, reflects intrinsic properties of the population, such as entrenched cultural values, social inertia, or loyalty to existing norms or products \cite{yildiz2013binary,mobilia2003does}. 
The dynamics of these competing forces provide a framework for studying tipping points, metastable states, and the long-term evolution of opinions in a wide variety of contexts.

In our model, the $n$ nodes of a network, the agents, support a binary opinion: the \textit{status quo} $r$ or the \textit{innovation} $b$.
As discussed previously, our model takes into account two competing biases at play. 
First, the \textit{disruptive bias} $p \in [0,1]$, which makes each agent perceive a neighbor's opinion as~$b$ with probability $p$, independently of the neighbor's actual opinion. 
This parameter models external interventions or the introduction of a new technology in the network. 
Second, the \textit{individuals' stubbornness} which amplifies the probability that the agents stick to opinion $r$, needing to observe multiple agents with opinion $b$ to switch.
This behavior of the dynamics captures the inherent resistance or attachment of the agents to their initial state.
In our dynamics, the agents sample $s$ random neighbors and adopt opinion $b$ only if the sampled neighbors are \textit{all} perceived as $b$ after accounting for the effect of the bias; otherwise, they adopt opinion~$r$.
Since the number of sampled neighbors $s$ is greater than one, the dynamics becomes nonlinear, generating complex phenomena and phase transitions, e.g., as for the 2-Choices dynamics \cite{cruciani2021phase} which has a closely related update rule where there is no stubbornness' effect.
On the other hand, if each node would sample one single neighbor, the dynamics would be equivalent to the (biased) voter model \cite{holley1975ergodic}, where each agent copies the opinion of that one randomly sampled neighbor. 

While the role of network structure in shaping opinion evolution is well acknowledged, most real-world networks are too complex and variable to allow for a clean theoretical analysis. This motivates the study of opinion dynamics on \emph{random graphs}, which serve as idealized yet flexible models of large-scale social and technological networks. 
The use of randomness enables one to abstract from the microscopic details of any particular network instance and instead focus on the influence of macroscopic structural features---such as degree distributions, sparsity, or directionality---on global dynamical behavior.
In this work we consider the \emph{Directed Configuration Model} (DCM)~\cite{van2024random}, that provides a natural setting to capture asymmetries in influence and information flow and to model systems where the ability to be influenced and the capacity to influence others are decoupled. This is crucial for representing many online platforms (e.g., X or Instagram), where users may have very different numbers of followers and followees.
Our aim in studying biased opinion dynamics on DCM random graphs is to uncover how simple, local update rules interact with global network heterogeneity and directionality. In this context, randomness is not a bug, but a feature: it allows us to identify the key statistical features of the degree sequences that govern macroscopic phenomena, such as phase transitions in consensus dynamics, while avoiding overfitting to any specific network topology.

\subsection{Our Contribution}

In many real-world social systems, individuals form opinions based on limited, potentially biased information about their peers. This is particularly evident in digital platforms, where users are exposed to content through noisy, algorithmically-curated feeds, often favoring certain narratives. Understanding how such local distortions aggregate into global consensus---or persistent disagreement---is a fundamental challenge in the study of opinion dynamics.

In this work, we introduce and analyze a novel model of \emph{biased majority dynamics} on directed graphs. Each node in the network holds a binary opinion---\textit{red} or \textit{blue}---and updates asynchronously by sampling the opinions of $s=2$ out-neighbors; in this presentation we fix the level of stubbornness $s$ for the sake of simplicity, but this is not restrictive as we discuss later in this section.
Specifically, at Poissonian times, a node selects two out-neighbors uniformly at random and observes each as blue with probability \( p \in [0,1] \), independently of their true state. 
The node adopts opinion blue if both observed neighbors appear blue, and otherwise chooses red. 
This setup models a basic form of \emph{disruptive bias}: even when red is the prevailing opinion, blue can spread if it is more visible or salient.

Despite the simplicity of the rule, the model exhibits rich and surprising behavior. While the system always eventually reaches the absorbing all-blue state when $p>0$, our focus is on the timescale and nature of this convergence. In particular, we investigate how long the red opinion persists when the dynamics starts from a uniform red configuration. Our main result shows that, when the underlying graph is drawn from the DCM, the system undergoes a sharp \emph{phase transition} as the bias parameter \( p \) varies (Theorem~\ref{theorem}). Below a critical threshold \( p_c \), a positive fraction of red opinions persists for a long time; above it, the system rapidly reaches consensus on the blue opinion.

A key element in our analysis is a duality between the opinion dynamics and a system of labeled particles---COalescing, BRAnching, and Dying (COBRAD) according to the same random environment (Proposition~\ref{prop:duality}). This dual system enables us to characterize the evolution of opinions via extinction properties of particles. Crucially, we find that the macroscopic behavior of the system depends only on two simple statistics of the degree sequence: the harmonic mean of the out-degrees weighted by in-degree (denoted \( \varrho \)), and its unweighted counterpart \( \lambda \) (see Eqs.~\eqref{eq:rho} and~\eqref{eq:hyp-rho}). These quantities capture how influence and exposure are distributed across the network and determine both the critical threshold \( p_c(\varrho) \) and the limiting red density \( q_\star(p, \varrho, \lambda) \) in the subcritical regime (resp.~Eqs.~\eqref{eq:pc} and~\eqref{eq:qstar}).
All our theoretical statements extend to arbitrary integer levels of stubbornness $s\ge 2$, resulting in higher-order (in $s$) degree statistics that influence the process quantitatively but not qualitatively.

Our results shed light on how structural asymmetries and local misperceptions can interact to sustain disagreement or drive consensus. Beyond the theoretical interest, this work provides a minimal framework to understand the dynamics of biased information spread in complex networks.

\section{Related Work}\label{sec:related}

The problem we address intersects multiple areas with related settings and diverse motivations. 
We focus on discussing the contributions most relevant to this paper.

\subsection{Opinion Diffusion and Consensus}

Opinion dynamics studies how groups of agents adjust their beliefs under the influence of peers and external factors. For a broad overview of opinion dynamics in multi-agent systems, see \cite{coates2018unified}.
The classical goal is to analyze the time it takes for the agents to reach a \textit{consensus} on some opinion.
A key distinction in the field lies in how opinions are represented. The case of continuous opinions has received significant attention, particularly in social sciences and economics \cite{degroot1974reaching,friedkin1990social,PROSKURNIKOV201765}. 
In this paper, however, we focus on the case of discrete opinions.

One classical model for discrete opinions is the \textit{majority dynamics}, where agents adopt the majority opinion among their neighbors. This rule was first studied in connection with agreement phenomena in spin systems \cite{krapivsky2003dynamics}. 
More recent studies address the role of network topology in the majority dynamics, showing under which conditions an initial majority can be overturned or not \cite{auletta2015minority,auletta2018reasoning,gartner2018majority}.

Another classical model is the \textit{voter model}, in which agents copy the opinion of a random neighbor. Originally inspired by studies of spatial conflicts in biology and interacting particle systems \cite{clifford1973model,holley1975ergodic,liggett2012interacting}, it has been extensively analyzed, with results that provide tight bounds on convergence times for various graph structures \cite{hassin1999distributed,cooper2013coalescing} and characterize the consensus probability on a given opinion \cite{donnelly1983finite}.

\subsection{Stubbornness and Bias in Opinion Dynamics}

A relevant area of study focuses on the impact of stubborn agents on opinion dynamics. \textit{Stubborn} agents are generally characterized by their inclination to favor either their initial or current opinion, while \textit{zealots} are defined as agents who permanently adhere to their initial opinion without any possibility of change. 
The impact of stubborn agents in the continuous-opinion dynamics has been considered for the main models \cite{wai2016,shirzadi2025do}, also considering external sources of bias \cite{out2024impact}.
Within the discrete-opinion framework, which this article examines, the influence of stubborn agents and zealots has been analyzed in the context of both the voter model \cite{mobilia2003does,mobilia2007role,10.1145/2538508} and the majority dynamics \cite{auletta2017information,mukhopadhyay2020voter}.
These studies have explored their ability to overturn an initial majority, shedding light on the conditions under which such a reversal is feasible.

Another line of work has analyzed the role of external bias toward a preferred opinion.
Only a handful of analytically rigorous studies have addressed the line of research of biased opinion dynamics. These include biased adaptations of the voter model \cite{sood2008voter,BerenbrinkGKM16,mukhopadhyay2020voter,anagnostopoulos2022biased,becchetti2023voter} and of other majority-based dynamics \cite{lesfari2022biased,cruciani2023phase,Mukhopadhyay_2024}, occasionally explored in the context of evolutionary game theory \cite{montanari2010spread}.
These studies explore various models, differing in how the bias is applied to update rule or in the temporal evolution of the process, e.g., synchronous vs asynchronous updates. It is worth noting that these factors can significantly influence the behavior of the resulting dynamics. In particular, as highlighted in several recent works \cite{hindersin2014counterintuitive,cooper2018discordant,anagnostopoulos2022biased}, even seemingly minor modifications to a model can lead to substantial differences in its behavior, making it difficult to predict whether the findings from one model will translate into another.

Other related works, have explored the impact of the influence exerted by small sets of \textit{elite} nodes \cite{10.1007/978-3-319-55471-6_7,10.1145/3288599.3288633,cruciani2021phase} and proposed to mitigate this influence by introducing random edges into the network topology \cite{OUT24}.
Some studies, instead, analyzed the effect of uniform \textit{noise} on opinion dynamics, rather than a bias toward a stronger/preferred opinion \cite{d2022phase,isa2022phase}.
Finally, the role of bias, stubborn agents, and social influence has been studied in collective decision making and election scenarios \cite{stewart2019information,BARA2022103773,CORO2022104940,faliszewski2022opinion}.

\subsection{The Role of Duality in Opinion Dynamics}
Duality has proven to be a powerful analytical tool in the study of interacting particle systems and, in particular, opinion dynamics on graphs. In its classical form, duality provides a correspondence between the original process and a simpler or more tractable auxiliary process evolving in a dual time direction. A prominent example is the \emph{voter model}, which is dual to a system of \emph{coalescing random walks} \cite{holley1975ergodic}. In this dual framework, the probability that a set of agents share a common opinion at time $t$ can be expressed as the probability that their associated random walks have coalesced by time $t$. This connection has been instrumental in deriving results on consensus times, fixation probabilities, and spatial correlations \cite{liggett2012interacting}.

Beyond the voter model, duality has also been extended to more complex systems, including models with asymmetric interaction rules or with branching mechanisms. For example, biased voter models and majority dynamics have been analyzed using dual processes involving coalescing-branching random walks \cite{cooper2016coalescing}. In these cases, duality provides not only computational convenience but also a probabilistic interpretation of how information or opinions propagate and interact over time. Our work contributes to the class of studies that exploit duality to analyze opinion dynamics, by relating the evolution of opinions to a tractable dual process that captures the flow of influence across the underlying graph, in a setting characterized by competing biases.

\section{Model and Results}\label{sec:model}

\begin{figure}[!h]
    \centering
    \includegraphics[width=0.7\linewidth]{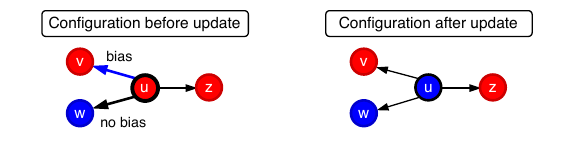}
    \caption{Graphical example of the update rule: vertex $u$ samples two neighbors $v,w$; the bias is applied on $v$, that is seen as \textit{blue}, but not on $w$. 
    Vertex $u$ updates to blue as it sees blue twice (effect of bias on $v$ and $w$ is blue).}
    \label{fig:update rule}
\end{figure}

Let $\cG=(\cV,\cE)$ be a \textit{directed graph} with $|\cV|=n$, and in which every vertex has out-degree at least $1$.
For a given vertex $x \in \cV$, we call $\cN_x=\{y\in\cV\,|\, (x,y)\in\cE\}$ the set of out-neighbors of~$x$. Each vertex is equipped with a binary opinion which we will identify as blue, $b$, or red, $r$. For $x\in\cV$ and $t\ge 0$ we write $c_x(t)\in\{b,r\}$ for the opinion of vertex $x$ at time $t$. We attach to each vertex an independent Poisson process of rate $1$. For a given vertex $x\in\cV$, at each arrival of the corresponding Poisson process the vertex selects $s=2$ (the \textit{stubbornness}) out-neighbors uniformly at random, with replacement.
The vertex then observes the color of such two neighbors: regardless of their actual colors, there is a probability $p\in[0,1]$ (the \emph{disruptive bias}) that the observed neighbor is seen blue. Each observation is independent of the others. After the observations the vertex assumes the color blue only if \emph{both} neighbors are seen blue, otherwise it assumes color red. See an example in Figure~\ref{fig:update rule}.

\begin{figure*}[!t]
    \centering
    \includegraphics[width=0.9\linewidth]{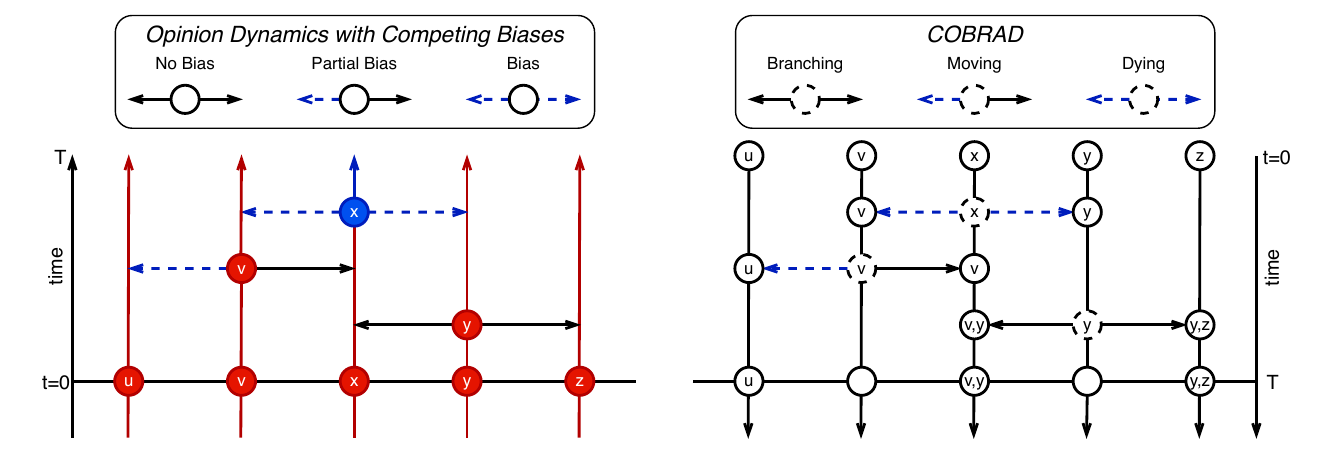}
    \caption{Graphical representation of the opinion dynamics with competing biases (left) and of COBRAD (right). On the left, in chronological order, $y$ samples $x,z$ without the effect of the bias, $v$ samples $u,x$ with the effect of the bias on $u$, and $x$ samples $v,y$ with the effect of the bias on both.
    On the right, in chronologically reverse order, the particle labeled $x$ dies, the particle labeled $v$ moves to vertex $x$, and the particle labeled $y$ branches to both vertices $x$ and $z$. At time $0$ in COBRAD there are no particles labeled $x$, hence vertex $x$ in the opinion dynamics with competing biases at time $T$ has color $b$.}
    \label{fig:cobrad}
\end{figure*}

Notice that the presence of the bias $p$ creates a \emph{frustration} in the system: on the one hand, each vertex has a tendency on adopting the opinion red, i.e., is \textit{stubborn}, since ties are broken in favor of this opinion; on the other hand, the presence of the \textit{bias} favors the opinion blue. It is clear that, regardless of the initial configuration, the system will be eventually absorbed in a configuration in which every vertex is blue. 
In what follows we will denote by $\xi_t\in\{r,b\}^{\cV}$ the opinion configuration at time $t\ge 0$, and use the shorthand notation $\xi^r$ (resp. $\xi^b$) to denote the configuration in which every vertex has opinion $r$ (resp. $b$). We will further write $R_t$ to denote the set of vertices sharing the opinion $r$ at time $t\ge 0$, and we will assume that the initial configuration is $\xi^r$. We are interested in understanding the time it takes for the system to reach the configuration $\xi^b$ as a function of the bias, $p$, and on the geometry of the underlying graph, $\cG$.

It will be convenient to describe our opinion dynamics by means of a so-called \emph{graphical construction}.  To each $x\in\cV$ we associate a marked Poisson process of rate 1. For each $x\in \cV$ and for each arrival $t\ge 0$ of the associated Poisson process we attach a mark 
\begin{equation}
    \begin{split}
(N_1(x,t),N_2(x,t),M_1(x,t),M_2(x,t))\\
\in\cN_x\times\cN_x\times\{0,1\}\times\{0,1\}\,.
    \end{split}
\end{equation} More precisely, the four components of the marks are sampled jointly independently, with the first two being marginally uniformly distributed and the last two being distributed as ${\rm Bern}(p)$.
$N_1(x,t)$ and $N_2(x,t)$ correspond to the neighbors of $x$ which are observed at time $t$, while the other two marks represent a bias in that observation: if $M_1(x,t)=1$ (resp. $M_2(x,t)=1$), then the neighbor $N_1(x,t)$ (resp. $N_2(x,t)$) is seen blue regardless of its current color. 
Then, the vertex $x$ becomes blue if both the sampled neighbors are seen blue. 
Otherwise, if $x$ sees at least a red neighbor, then it will become red.

\subsection{The Dual Particle System}\label{sec:dual-ips}
A key observation is that the opinion dynamics described above is dual to a system of COalescing, BRAnching and Dying labeled particles, which we will denote by COBRAD. 
At time $0$ each particle sits on a vertex, and it is labeled by the label of such a vertex. 
Moreover, when two particles with the same label sit on the same vertex, we will treat them as a single particle. 
Therefore, the process $(\eta_t)_{t\ge 0}$ has state space $\cV^\cV$. 
In words, for every $x\in \cV$, $\eta_t(x)\subset\cV$ denotes the set of the labels of the particles sitting at $x$ at time $t$. 
Moreover, for any $y\in \cV$ we will write $y\in \eta_t$ to denote that there exists some $x\in\cV$ such that $y\in \eta_t(x)$. 
The process evolves by means of the same marked Poisson processes used for the opinion dynamics. 
In particular, at the arrivals of the Poisson process associated to vertex $x$, one of the following scenarios occur:
\begin{itemize}
    \item If $M_1(x,t)=M_2(x,t)=0$, then each particle at $x$ undergoes a branching, i.e., it is removed from $x$ and it produces two particles (with the same label) which are placed at $N_1(x,t)$ and at $N_2(x,t)$.
    \item If $M_1(x,t)=0$ while $M_2(x,t)=1$, then each particle at $x$ moves to $N_1(x,t)$.
    \item Similarly, if $M_2(x,t)=0$ while $M_1(x,t)=1$, then each particle at $x$ moves to $N_2(x,t)$.
    \item Finally, if $M_1(x,t)=M_2(x,t)=1$ all the particles sitting at $x$ die.
\end{itemize}
Notice that, regardless of the initial configuration, the particles system will undergo an extinction in finite time \textit{almost surely}, i.e., with probability 1.

For a given directed graph, and for a prescribed time horizon $T>0$ we sample the marked Poisson processes that will be used to construct the opinion dynamics up to time $T$. Then, we consider the COBRAD particle system which uses the same marked Poisson process, but backward in time. More precisely, for some $A\subseteq \cV$ we start with the configuration $\eta_0=\eta^A$ having at site $x\in\cV$ a single particle (with label $x$) if and only if $x\in A$, and otherwise $x$ is empty at time 0.  
We consider the marked Poisson process 
\[
    (\widehat{\rm PP}_t(x))_{x\in \cV,t\in[0,T]}=({\rm PP}_{T-t}(x))_{x\in \cV,t\in[0,T]}.
\]
The following fact is an immediate consequence of the latter construction.
\begin{Proposition}\label{prop:duality}
For any $A\subseteq V$, $C\subseteq A$ and $T>0$
\begin{equation}
    \begin{split}
  \mathbf{P}\left(\cap_{x\in C}\{c_x(t)=b\}
  \cap_{x\in A\setminus C}\{c_x(t)=r\} \mid \xi_0=\xi^r\right)\\
  =\mathbf{P}\left(\cap_{x\in C}\{x\not\in \eta_t\}
  \cap_{x\in A\setminus C}\{x\in\eta_t \}\mid \eta_0=\eta^A\right)\,.
    \end{split}
\end{equation}
\end{Proposition}
In other words, a given subset of vertices $C\subseteq A$ has opinion $b$ at time $t$ if and only if in the associated particle systems the labels in $C$ are extinct at time $t$. See Figure~\ref{fig:cobrad} for a visual representation.

\subsection{The Directed Configuration Model (DCM)}\label{sec:DCM}
Despite the fact that the opinion dynamics described above and its dual particle system are well defined on any directed (or undirected) graph, in this work we focus on a benchmark model known as \emph{Directed Configuration Model}~\cite{van2024random}. Similarly to its undirected counterpart, \cite{bollobas1980probabilistic,van2024random}, such a random digraph model is defined as follows: for every fixed graph size, $|\cV|=n$, the in- and out-degree sequences are prescribed as parameters of the model, $(d_x^-,d_x^+)_{x\in\cV}$. We will be interested in the large graph limit, $n\to\infty$, so the asymptotic notation is referring to such a limit and the explicit dependence on $n$ will be dropped from the notation. We will further assume that such degree sequences satisfy the following requirements
\begin{equation}\label{eq:hyp}
\begin{cases}
 \sum_{x\in \cV}d_x^-=\sum_{x\in \cV}d_x^+=:m;
\\ \min_{x\in V} d_x^+\geq 2;
\\\Delta:=\max_{x\in V} d_x^+=O(1)\,;
\\\frac1n\sum_{x\in \cV}(d_x^-)^{2+\delta}=O(1)\qquad\text{for some $\delta>0$}.
\end{cases}
\end{equation}
Before commenting on these requirements, let us explain how to obtain a random graph out of the choice of such parameters: we attach to each vertex, $x\in\cV$, $d_x^-$ labeled \emph{tails} and $d_x^+$ labeled \emph{heads} and then take a uniformly random bijection from the set of tails to the set of heads. The first requirement is needed to guarantee that such a random bijection is well defined and it can be projected into a (multi-)digraph%
\footnote{We will in fact assume that multiple edges with the same source-destination are allowed, as well as self-loops. Despite that, we will omit the prefix \emph{multi-}.}. 
The second requirement guarantees that, with high probability, the digraph obtained by this procedure admits a unique strongly-connected component, see \cite{BCS2018,CaiPerarnau2023,CCPQ}. 
The last two requirements are of technical nature; nevertheless they are quite reasonable for a model of a real-world social network: thinking, e.g., at a social network like \emph{Instagram}, the number of people you follow is somehow constrained to be a bounded quantity, while it is natural to assume that there are few \emph{influencers} who have an enormous number of \emph{followers}. 
In a sense, the DCM serves as a benchmark for digraphs with given degree sequences, since it can be thought of as the \emph{maximum entropy ensemble} for graphs with such in- and out-degrees. Despite the size, $2n$, of the model parameter, it is known that the behavior of several diffusion models and opinion dynamics on these random graphs actually depends only on a few simple statistics of the degree sequences, see \cite{BCS2018,Avena,CQ20}. In the setup of this work, we aim at understanding which are the features of the degree sequence that have an impact on the behavior of our biased opinion dynamics. Perhaps surprisingly, we will show that the behavior is in fact dictated only by the following statistics:
\begin{equation}\label{eq:rho}
\begin{aligned}
&\varrho_n:=\sum_{x\in\cV}\frac{d_x^-}{m}\frac{1}{d_x^+}\,,\quad\lambda_n:=\frac1n\sum_{x\in\cV}\frac1{d_x^+}\,,
\end{aligned}
\end{equation}
namely, $\varrho$ is the average of the inverse out-degree of a vertex sampled with probability proportional to its in-degree; while $\lambda$ is the average of the same quantity but with respect to the uniform distribution. Notice that, thanks to \eqref{eq:hyp} we have $\rho,\lambda\in(0,\frac12]$.
It is worth to point out that the key role of the statistics $\varrho$ already appeared in other works investigating stochastic processes on the DCM \cite{BCS2018,Avena}, and has also been related to the spectral properties of such random graph \cite{Coste}. 
Clearly, the quantities in \eqref{eq:rho} depend on $n$, therefore, in order to avoid inconsistencies, we will assume that the (sequence of) degree sequences is such that
\begin{equation}\label{eq:hyp-rho}
\lim_{n\to\infty}\varrho_n=\varrho\,,\qquad \lim_{n\to\infty}\lambda_n=\lambda\,.
\end{equation}
 
\subsection{Results}\label{sec:results}
Our main result is the identification of a phase transition on the behavior of the biased opinion dynamics described above when taking place on the DCM random graph. To phrase our theorem let us start with a definition.
\begin{Definition}
    We will call \emph{slowly diverging sequence} a positive sequence $(f_n)_{n\in\mathbb{N}}$ such that
    \begin{equation}
    \liminf_{n\to\infty} f_n=+\infty\,,\qquad \limsup_{n\to\infty}\frac{f_n}{\log (n)}=0\,.
    \end{equation}
\end{Definition}
\begin{theorem}\label{theorem}
Consider, for all $n\in\mathbb{N}$, the degree sequences $(d_x^-,d_x^+)_{x\in\cV}$ satisfying \eqref{eq:hyp} and \eqref{eq:hyp-rho} and call 
\begin{align}
p_c(\varrho)&:=\frac{\sqrt{1-\varrho}-(1-\varrho)}{\varrho}\,,
\label{eq:pc}
\\
q_\star(p,\varrho,\lambda)&:=
\left(1-\tfrac{p^2}{(1-p)^2(1-\varrho)} \right)\left(1-\tfrac{p^2(\lambda-\varrho)}{1-\varrho} \right)\,.
\label{eq:qstar}
\end{align}
Take $(t_n)_{n\in\mathbb{N}}$ an arbitrary slowly diverging sequence and call $\mathbf{P}$ the joint law of the DCM and the opinion dynamics. Then:
\begin{itemize}
    \item if $p\ge p_c$ then for all $\varepsilon>0$
    \[
    \lim_{n\to\infty}\mathbf{P}\left(\frac{|R_{t_n}|}{n}<\varepsilon\,\bigg\rvert\, \xi_0=\xi^r \right)=1\,;
    \]
     \item if $p< p_c$ then for all $\varepsilon>0$ 
     \[
     \lim_{n\to\infty}\mathbf{P}\left(\left|\frac{|R_{t_n}|}{n}-q_\star\right|<\varepsilon \,\bigg\rvert\, \xi_0=\xi^r\right)=1\,. 
     \]
\end{itemize}
\end{theorem}
\begin{Remark}
It is crucial to note that $q_\star(p,\varrho,\lambda)\in[0,1)$ for all $p<p_c(\varrho)$ and $\lambda,\varrho\in(0,1/2]$, and that the function $p\mapsto q_\star(p,\varrho,\lambda)$ is decreasing regardless of $\lambda,\varrho\in(0,1/2]$.
\end{Remark}
\subsection{Interpretation of the Results}\label{sec:interpretation}

Our main theorem reveals that the behavior of the biased opinion dynamics undergoes a \emph{sharp phase transition} as a function of the bias parameter \( p \), with a critical threshold \( p_c(\varrho) \) that depends on the structure of the underlying directed graph. Below this threshold, a persistent density of red opinions survives for a long time; above it, the system rapidly converges to full consensus on blue---even if all individuals initially hold the red opinion.

\begin{figure}[!th]
    \centering
\includegraphics[width=0.5\linewidth]{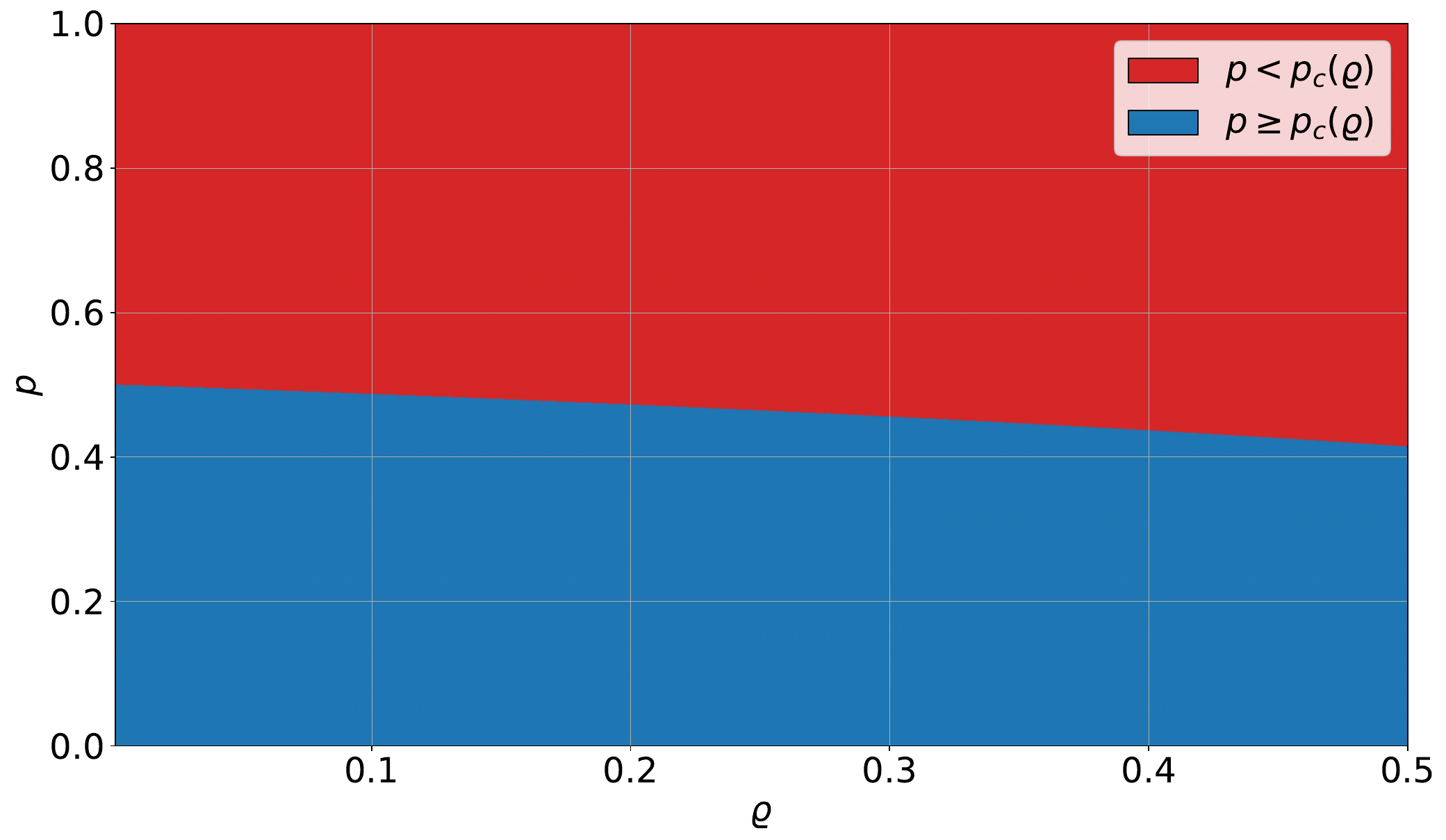}
    \caption{On the $x$-axis have the value of $\varrho$, while on the $y$-axis the value of $p$. The blue (resp. red) area represents couples $(\varrho,p)$ for which we have $p> p_c(\varrho)$ (resp. $p< p_c(\varrho)$).}
    \label{fig:survival}
\end{figure}

This result offers a number of conceptual insights.
First, it highlights the \emph{nontrivial interplay between bias and network structure}. The fact that the critical bias depends on the structural parameter \( \varrho \)---which reflects how in-degrees correlate with out-degrees---indicates that the same level of local misperception may have dramatically different global effects depending on how influence and exposure are distributed across the network. In particular, the system is more susceptible to blue consensus (i.e., has lower \( p_c \)) when individuals with many incoming connections tend to have fewer outgoing ones (i.e., has larger $\varrho$), cf. Figure~\ref{fig:survival}.

Second, the limiting fraction of red individuals in the subcritical regime is given explicitly by the quantity \( q_\star(p,\varrho,\lambda) \), showing that \emph{partial consensus (or polarization)} is not a transient phenomenon but a stable long-time outcome under certain conditions. In other words, \emph{competing biases alone are sufficient to sustain diversity of opinions}.

\begin{figure}[!th]
    \centering
    \includegraphics[width=0.49\linewidth]{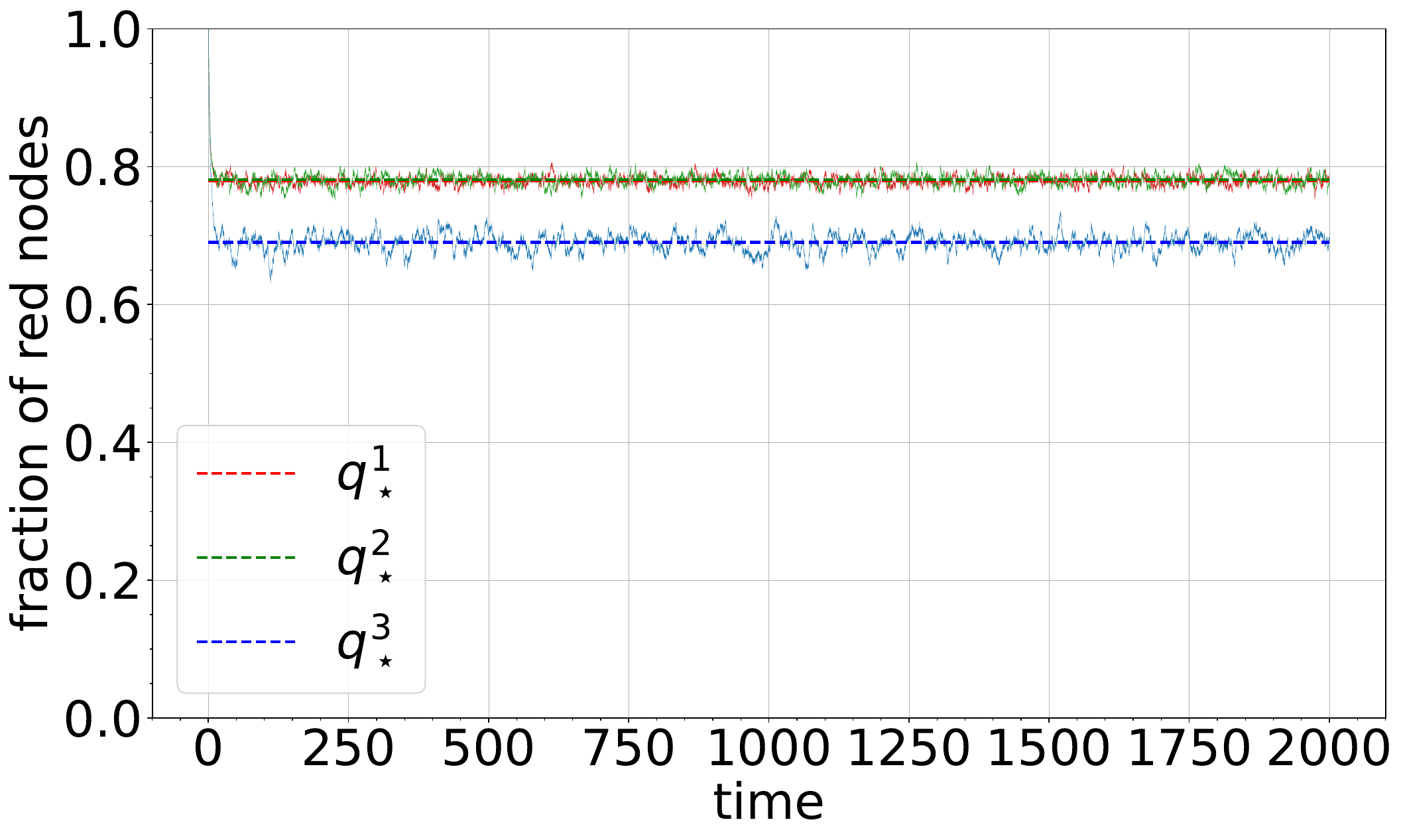}
    \includegraphics[width=0.49\linewidth]{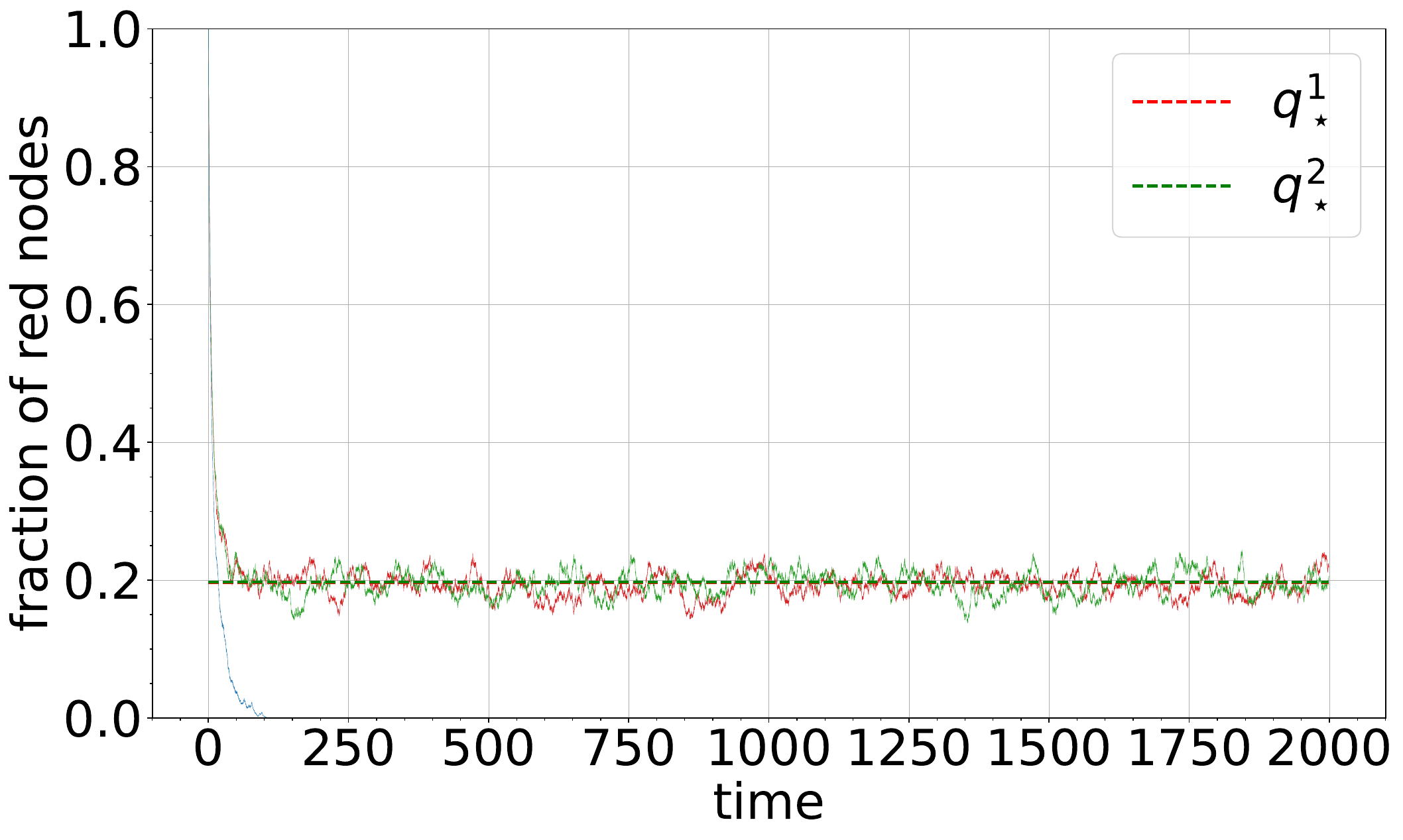}
    \caption{We run the biased opinion dynamics on three DCM graphs with $n=10^4$ vertices: in the first all vertices have in- and out-degrees equal to $6$ (red); in the second, half of the vertices have in-degree $10$ and out-degree $5$, and the other way around for the other half (green); in the third, half of the vertices have in-degree $10$ and out-degree $2$, and vice versa for the second half (blue). Notice that the red and the green graph share the same value of $\varrho=\frac{1}{6}$ (hence, they have the same $p_c\approx 0.477$) while the red and the blue graph have the same number of edges (but in the blue case $\rho=\frac{13}{30}>\frac{1}{6}$). Notice also that the value of $\lambda$ is $\frac16$, $\frac3{20}$ and $\frac3{10}$, in the blue, red and green case, respectively. The value of the critical threshold in the case of the blue graph is $p_c\approx 0.429$. On the left a simulation for $p=0.3$, which is subcritical in all the three cases; on the right a simulation for $p=0.45$, which is subcritical for the red and the green graphs, but supercritical for the blue one. The dashed lines indicate the theoretical prediction of the long-term density of red opinions, i.e., $q_\star(p,\varrho,\lambda)$. Notice that in the case $p=0.3$ (left) we have $q_\star(\frac3{10},\frac16,\frac16)\approx 0.7796$ (red), $q_\star(\frac3{10},\frac16,\frac3{20})\approx 0.7810$ (green) and $q_\star(\frac3{10},\frac16,\frac3{10})\approx 0.6902$ (blue); while in the case $p=0.45$ (right) we have $q_\star(\frac9{20},\frac16,\frac1{6})\approx 0.1967$ (red) and $q_\star(\frac9{20},\frac16,\frac3{20})\approx 0.1975$ (green).}
    \label{fig:simulazioni}
\end{figure}

Third, and perhaps most surprisingly, the critical threshold and the limiting behavior of the system depend only on \emph{aggregated statistics} \( \varrho \) and \( \lambda \), rather than on the full degree sequence. This universality suggests that the \emph{global behavior is insensitive to microscopic details}, and instead dictated by a coarse balance between influence and susceptibility. From a modeling perspective, this greatly simplifies the understanding of consensus dynamics on large directed graphs: if one is convinced that the structure of the network in analysis is dictated only by its first-order properties (i.e., the degrees), then one can reason effectively in terms of just two scalar quantities.
This can be observed in Figure~\ref{fig:simulazioni}.

Finally, from a broader viewpoint, our findings emphasize how \emph{individual-level misperceptions} (i.e., biased observations) can amplify and propagate across the network in a highly nonlinear way. Even a small probability of misinterpreting neighbors' opinions can lead to a tipping point where the majority flips. This provides a potential explanation for the emergence of dominant narratives or widespread misinformation in digital platforms, where the structure of influence is asymmetric and noisy observations are the norm.

\section{The Associated Random Tree}\label{sec:tree}
At the core of our analysis stands the following intuition: it is well-known that sparse random graphs can be locally approximated by random trees, whose law depends on the underlying degree sequence. For the precise mathematical formulation of this fact we refer to \cite{van2024random} and, more specifically to \cite{BCS2018} for the case of the DCM. 
In this section we focus on the analysis of the COBRAD process on such approximating random trees, showing how the key quantities $p_c$ and $q_\star(p)$ appear in the analysis. 
In the supplemental material, which is omitted for the sake of space, we show how to translate the analysis in this section on a result for the random graph model. 
Therefore, we do not aim at proving here in full details the result in Theorem~\ref{theorem}, but rather on providing an intuition for its validity.

The key idea can be summarized as follows. Let us focus on a given vertex $x\in\cV$. Thanks to the duality with the COBRAD process we know that the probability that $c_x(t)=r$ coincides with the probability that the COBRAD process initialized with a single particle at $x$ is not extinct at time~$t$. 
Now, if we approximate the local out-neighborhood of $x$ by a random tree, we can approximate the probability of survival of the COBRAD process at time $t$ on the graph with the survival of the same process on such an approximating random tree.

Fix $n\ge 1$ and consider the in- and out-degree sequences $(d_x^-,d_x^+)_{x\in\cV}$. We construct the approximating random directed tree as follows: the root has a deterministic number of children, $d_{\rm root}$, while each other vertex on the tree has a number of children, $X$, such that the probability that $X=k$ corresponds to the probability that, sampling a vertex in $\cV$ with probability proportional to its in-degree, we obtain a vertex with out-degree $k$. 
In other words, called $P_n$ the law of the offspring distribution, we have
\begin{equation}\label{eq:xidist} P_n(X=k)=\sum_{x\in\cV}\frac{d_x^-}{m}\mathds{1}_{\{d_x^+=k\}}\,,\qquad k\in[2,\Delta]\,.
\end{equation}
Once the tree has been constructed, we consider the COBRAD process starting with a single particle at the root. 
Being the tree directed, the particles can only perform downward transitions along the tree. 
In particular, if a particle ever visits a certain vertex of the tree having degree $d$, then such a particle will eventually:
\begin{itemize}
    \item Die with probability 
    \[
    p_0(d)=p^2\,;
    \]
    \item Move to one of the children with probability
    \[
    p_1(d)=(1-p)\left[2\left(1-\frac1d\right)p+\frac1d\left(1+p\right)    \right]\,;
    \]
    \item Branch into two new particles going on two different children of the current vertex with probability
    \[
    p_2(d)=\left(1-\frac{1}{d}\right)(1-p)^2\,.
    \]
\end{itemize}

This process can therefore be seen as a \emph{percolation} on the random tree, where each edge is retained if and only if a particle eventually crosses the edge. Called $\mathbb{P}_n$ the joint law of the tree construction and of the COBRAD, we can reinterpret the percolated tree as a random subtree in which each vertex can only have $0,1$ or $2$ children. 
For the root, the probability of each of these outcomes is given by $p_0(d_{\rm root})$, $p_1(d_{\rm root})$ and $p_2(d_{\rm root})$, while for all the other vertices such probabilities are given by
\begin{equation*}
\begin{split}
p_0&=\sum_{x\in\cV}\frac{d_x^-}{m}p_0(d^+_x)=p^2\,,\\
p_1&=\sum_{x\in\cV}\frac{d_x^-}{m}p_1(d^+_x)=(1-p)(2p-p\varrho_n+\varrho_n)\,,\\
p_2&=\sum_{x\in[n]}\frac{d_x^-}{m}p_2(d_x^+)=(1-\varrho_n)(1-p)^2\,.
\end{split}
\end{equation*}
We are now interested in evaluating the probability that such a percolated tree is infinite. 
Given the special role of the root, we start by looking at the probability that the percolated tree starting at one of the root's children is infinite (so that every vertex on this subtree has the same offspring distribution). 
By the classical theory of random trees (see, e.g., \cite{LyonsPeres}) one has that the probability that such tree is infinite is zero as soon as the expected number of children is smaller than or equal to 1, i.e.,
\begin{equation}
    p_1+2p_2\le 1\iff p\ge  p_c(\varrho_n)\,.
\end{equation}
More generally, the probability that such a tree is finite can be obtained by looking at the smallest solution of the equation
\begin{equation}\label{eq:sol}
p_0+p_1z+p_2z^2=z\,,
\end{equation}
that is
\begin{equation}
z_\star(p,\varrho_n)=\frac{p^2}{(1-p)^2(1-\varrho_n)}\wedge 1\,.
\end{equation}
Note that the above minimum is $1$ if and only if $p\ge p_c(\varrho_n)$.

Keeping into account the role of the root, we have that the percolated tree starting at the root is finite with probability
\begin{equation}\label{eq:zeta}
\begin{split}
\hat{z}_\star(d_{\rm root},p,\varrho_n)=\,&p_0(d_{\rm root})+p_1(d_{\rm root})z_\star(p,\varrho_n)\\
&\qquad\quad\,\,\,\,+p_2(d_{\rm root})z_\star(p,\varrho_n)^2\,.
\end{split}
\end{equation}
At this point, using such an approximation, the density of red vertices at time $t$ is given by adding up the probabilities in \eqref{eq:zeta} when the root is each $x\in\cV$. As a consequence:
\begin{equation}
\frac{|R_t|}{n}\approx \frac1n\sum_{x\in\cV}\big(1-\hat z_\star(d_x^+,p,\varrho_n)\big)=q_\star(p,\varrho_n,\lambda_n)\,,
\end{equation}
where the latter quantity is defined as in \eqref{eq:qstar}.

\section{Conclusion and Future Work}

We have introduced a simple yet expressive model of opinion dynamics with competing biases and shown that exhibits a sharp phase transition driven by the strength of the disruptive bias. Our analysis on the Directed Configuration Model reveals that long-term disagreement (or rapid consensus) can emerge purely from the interplay between local misperceptions and global network structure.

The duality with a particle system allowed us to characterize the metastable behavior of the process and to identify the critical threshold \( p_c(\varrho) \) and limiting red density \( q_\star(p, \varrho, \lambda) \) in the subcritical regime. Remarkably, both quantities depend only on coarse statistics of the degree sequence, suggesting a form of universality in the large-network limit.

These findings highlight the power of minimal stochastic models in capturing essential features of social dynamics under noise and structural asymmetry.

On the mathematical side, an intriguing open problem is to understand the precise time to consensus in the large-graph limit, particularly in relation to metastability and sharp transition windows. In the spirit of the mathematical literature on the contact process on random graphs \cite{val1,val2}, we formulate the following conjecture:

\begin{Conjecture}In the same setup as Theorem \ref{theorem}
\begin{itemize}
    \item if $p\ge p_c$ then there exists some $\bar c(p)>0$ such that 
    \[
        \lim_{n\to\infty}\mathbf{P}\Big(R_{\bar c\log(n)}=\emptyset\,\mid\, \xi_0=\xi^r \Big)=1\,;
    \]
     \item if $p< p_c$ then there exists some $\underline c(p)>0$ such that  
     \[
     \lim_{n\to\infty}\mathbf{P}\Big(R_{e^{\underline c n}}\neq\emptyset\,\mid\, \xi_0=\xi^r\Big)=1\,. 
     \]
Moreover, called $\tau_b=\inf\{t\ge 0\mid \xi_t=\xi^b\}$,  
\[
\lim_{n\to\infty}\mathbf{P}\Big( \tau_b > t \cdot \mathbf{E}[\tau_b]\,\mid\, \xi_0=\xi^r\Big)=e^{-t}\,.
\]
\end{itemize}
\end{Conjecture}

\appendix
\section{Appendix}\label{apx}
It will be fundamental in the proof of Theorem \ref{theorem} to consider the marginal probability distributions associated to the generation of the graph and the one associated to the opinion dynamics/particle system on the graph. For this reason, we will write $\mathbf{P}^{\rm DCM}$ for the probability law of the graph generation and, for a given realization of the random digraph $G$, $\mathbf{P}^{\rm op}(\cdot\mid G)$ denotes the law of the associated  opinion dynamics (and, by duality, of the COBRAD dynamics) on $G$. To improve readability we will drop the conditioning on $G$ from the notation, but we will in general consider $\mathbf{P}^{\rm op}(\cdot)$ to be a random variable with respect to the probability distribution $\mathbf{P}^{\rm DCM}$.

Our proof will exploit some recent result concerning the local structure of a DCM digraph and the corresponding technical insights coming from these works (see \cite{BCS2018,CCPQ,Avena}).

The first ingredient is the above-mentioned local approximation of the out-neighborhood of a given vertex $x\in \cV$. In \cite[Sec. 2.2.2]{CCPQ} it is shown that, for an arbitrary slowly diverging sequence, $h_n\in\bbN$, the out-neighborhood of $x$ of depth $\hslash=\gamma \log_\Delta(n)$ can be coupled with the random tree introduced in Section~\ref{sec:tree} with high probability as soon as $\gamma$ is sufficiently small. Moreover, arguing as in \cite[Sec. 3]{CCPQ} (or, similarly, in \cite[Sec. 6]{Avena}) one can average with respect to $\mathbf{E}^{\rm DCM}$ the probability $\mathbf{P}^{\rm op}(c_x(t)=r)$ by generating the graph locally around $x$ together with the COBRAD process, up to the first time in which a particle of the COBRAD process exits the neighborhood of $x$ of depth $\hslash$. Since the out-neighborhood of any vertex up to depth $\hslash$ has $O(\Delta^\hslash)$ vertices, taking $t$ slowly diverging, by a union bound it follows that the probability that a COBRAD particle exits the out-neighborhood within time $t$ can bounded by
\begin{equation}
\Delta^\hslash \mathbb{P}({\rm Poi}(t)>\hslash)=e^{-\Omega(\hslash\log(\hslash/t))}=o(1)\,,
\end{equation}
where we used classical bounds on the large deviations of Poisson random variables and the fact that $t$ is slowly diverging.

\subsection{Proof of the supercritical case: $p\ge p_c(\varrho)$}
Fix $\varepsilon>0$ and let $t=t_n$ slowly diverging. Fix the degree sequence $(d_x^+,d_x^-)_{x\in\cV}$ satisfying \eqref{eq:hyp}-\eqref{eq:hyp-rho} and $p\ge p_c(\varrho)$. We will assume that $\xi_0=\xi^r$ and we drop this information from our notation. We aim at showing that
\begin{equation}\label{eq:16}
\lim_{n\to\infty}\mathbf{P}\left(|R_t|>\varepsilon n  \right)=0\,,
\end{equation}
namely, for any $\delta>0$
\begin{equation}
\lim_{n\to\infty}\mathbf{P}^{\rm DCM}\left(\mathbf{P}^{\rm op}(|R_t|>\varepsilon n )>\delta \right)=0\,.
\end{equation}
By Markov inequality we have,
\begin{equation*}
\mathbf{P}^{\rm op}\left(|R_t|>\varepsilon n \right)\le \frac{\mathbf{E}^{\rm op}[|R_t|]}{\varepsilon n}\,,\quad \mathbf{P}^{\rm DCM}\text{-a.s.}
\end{equation*}
and therefore, it is enough to prove that, $\forall\delta>0$
\begin{equation}
\lim_{n\to\infty}\mathbf{P}^{\rm DCM}\left(\mathbf{E}^{\rm op}\left[|R_t|\right]>\delta n \right)=0\,,
\end{equation}
and the latter, again by Markov inequality, is implied by
\begin{equation}\label{eq:19}
\lim_{n\to\infty}\frac{\mathbf{E}^{\rm DCM}\left[\mathbf{E}^{\rm op}[|R_t|]\right]}{n}=\lim_{n\to\infty}\frac{\mathbf{E}\left[|R_t|\right]}{n}=0\,.
\end{equation}
Notice that
\begin{equation}
\mathbf{E}^{\rm op}[|R_t|]=\sum_{x\in \cV}\mathbf{P}^{\rm op}(c_t(x)=r)\,,
\end{equation}
and therefore
\begin{equation}\label{eq:21}
\mathbf{E}[|R_t|]=\sum_{x\in \cV}\mathbf{E}^{\rm DCM}\big[\mathbf{P}^{\rm op}(c_x(t)=r)\big]\,.
\end{equation}
Now, for a given $x\in\cV$ call $\cC_x$ the event that the coupling of the $\hslash$-neighborhood of $x$ and the random tree is successful and call $\cD_x$ the event that by time $t$ no COBRAD particle exits the such $\hslash$-neighborhood. Then
\begin{equation}\label{eq:22bis}
\begin{split}
&\mathbf{E}^{\rm DCM}\big[\mathbf{P}^{\rm op}(c_x(t)=r)\big]\le \mathbf{P}^{\rm DCM}(\cC_x^c)+\mathbf{E}^{\rm DCM}[\mathbf{P}^{\rm op}(\cD^c_x)]+\mathbf{P}^{\rm tree}(\eta_t\neq \emptyset)\,,
\end{split}
\end{equation}
where $\mathbf{P}^{\rm tree}$ is the probability distribution of the COBRAD process on the random tree as described in the dedicated section.
In words, in order to have the COBRAD process alive at time $t$ starting from a unique particle at $x$ either:
\begin{itemize}
    \item The $\hslash$-neighborhood of $x$ is not coupled with the random tree;
    \item or a COBRAD particle exits the $\hslash$-neighborhood before time $t$;
    \item or, the COBRAD process on the random tree survives up to time $t$.
\end{itemize}
The three contributions all go to zero by the argument at the beginning of the appendix (and, for the third, in the section about the approximating random tree). In conclusion, each summand in \eqref{eq:21} is $o(1)$, from which we deduce the validity of \eqref{eq:19} and, in turns, of \eqref{eq:16}.

\subsection{Proof of the subcritical case: $p<p_c(\varrho)$}
Fix $\varepsilon>0$ and let $t=t_n$ slowly diverging. Fix the degree sequence $(d_x^+,d_x^-)_{x\in\cV}$ satisfying \eqref{eq:hyp}-\eqref{eq:hyp-rho} and $p< p_c(\varrho)$. To simplify the notation let $q_\star$ be $q_\star(p,\rho,\lambda)$. We will assume again that $\xi_0=\xi^r$ and aim at showing that
\begin{equation}\label{eq:22}
\lim_{n\to\infty}\mathbf{P}\left(\big||R_t|-q_\star n\big|>\varepsilon n  \right)=0\,,\qquad \text{in $\mathbf{P}^{\rm DCM}$}
\end{equation}
namely, for any $\delta>0$
\begin{equation}\label{eq:24}
\lim_{n\to\infty}\mathbf{P}^{\rm DCM}\left(\mathbf{P}^{\rm op}\big(\big||R_t|-q_\star n\big|>\varepsilon n \big)>\delta \right)=0\,.
\end{equation}
Let us start by calling
$$Y=\mathbf{E}^{\rm op}[|R_t|]\,,\qquad Z=\mathbf{E}^{\rm op}[|R_t|^2]\,,$$
which are both random variables with respect to the generation of the graph.
We first claim that
\begin{equation}\label{eq:claim1}
\mathbf{P}^{\rm DCM}\left( |Y-q_\star n|=o(n)\right)=1-o(1)\,,
\end{equation}
and, secondly,
\begin{equation}\label{eq:claim2}
\mathbf{E}^{\rm DCM}[Z]=(1+o(1))\mathbf{E}^{\rm DCM}[Y^2]\,.
\end{equation}
Consider now 
$$X=\mathbf{P}^{\rm op}\big(\big||R_t|-q_\star n\big|>\varepsilon n \big) $$
and notice that \eqref{eq:24} is equivalent to
$$\lim_{n\to\infty}\mathbf{P}^{\rm DCM}(X>\delta)=0\,,\qquad \forall\delta>0\,,$$
which is in turn implied (thanks to Markov inequality) by the estimate
\begin{equation}\label{eq:27}
\lim_{n\to\infty}\mathbf{E}^{\rm DCM}[X]=0\,.
\end{equation}
Notice that
\begin{equation*}
    \begin{split}
X&\le \mathbf{P}^{\rm op}\left(
\big||R_t|-Y\big|+|Y-q_\star n|>\delta n\right)\\
&\le\mathbf{P}^{\rm op}\left(
\big||R_t|-Y\big|>\frac\delta2 n\right)+\mathbf{1}\big(|Y-q_\star n|>\tfrac\delta2 n\big)\\
&\le \frac{4(Z-Y^2)}{\delta^2 n^2}+\mathbf{1}\big(|Y-q_\star n|>\tfrac\delta2 n\big)
    \end{split}
\end{equation*}
where in the last step we have used Chebyshev inequality.
Hence, taking the expectation,
\begin{equation*}
\begin{split}
    \mathbf{E}^{\rm DCM}[X]&\le \frac{4 \mathbf{E}^{\rm DCM}[Z-Y^2]}{\delta^2 n^2}\\
    &\quad+\mathbf{P}^{\rm DCM}\big(|Y-q_\star(p)n|>\tfrac\delta2 n\big)= o(1)\,,
    \end{split}
\end{equation*}
where the latter inequality follows from \eqref{eq:claim1} and \eqref{eq:claim2}. This proves \eqref{eq:27} and, in turns, the desired result. Therefore, we are left to prove \eqref{eq:claim1} and \eqref{eq:claim2}.

\begin{proof}[Proof of Eq.~\eqref{eq:claim1}]
Let us start by computing the expectation of $Y$,
\begin{equation}
    \mathbf{E}^{\rm DCM}[Y]=\sum_{x\in\cV}\mathbf{E}^{\rm DCM}[\mathbf{P}^{\rm op}(c_x(t)=r)]\,.
\end{equation}
Arguing as in \eqref{eq:22bis} we have
\begin{equation}
 \mathbf{E}^{\rm DCM}[Y]=q_\star n +o(n)\,.
\end{equation}
Now, computing the second moment of $Y$ we get
\begin{equation*}
\begin{split}
\mathbf{E}^{\rm DCM}[Y^2]&=\sum_{x,y\in\cV}\mathbf{E}^{\rm DCM}[\mathbf{P}^{\rm op}(c_x(t)=r)\mathbf{P}^{\rm op}(c_y(t)=r)]\\
&=\sum_{x,y\in\cV}\mathbf{E}^{\rm DCM}[\mathbf{P}^{\rm op}_x(\eta_t\neq\emptyset)\mathbf{P}_y^{\rm op}(\eta_t\neq \emptyset)]
\end{split}
\end{equation*}
where we adopted the notation $\mathbf{P}^{\rm op}_x$ to denote the law of the COBRAD process started with a single particle at $x$. Now, it is enough to realize that, for a given couple $x,y$ the probability that the two out-neighborhoods have an overlap is $o(1)$ (thanks to \cite[Eq. (2.2)]{CCPQ}), therefore one can replace the expectation of the product of the two probabilities with the product of the expectations, hence concluding that
\begin{equation}\label{eq:30}
\mathbf{E}^{\rm DCM}[Y^2]=(1+o(1))\mathbf{E}^{\rm DCM}[Y]^2\,,
\end{equation}
and therefore the asymptotic estimate in \eqref{eq:claim1} follows by Chebyshev inequality.
\end{proof}
\begin{proof}[Proof of Eq.~\eqref{eq:claim2}]
    From the previous proof we already know that
    \begin{equation}
        \mathbf{E}^{\rm DCM}[Y^2]=(1+o(1))q_\star^2 n^2\,.
    \end{equation}
    Therefore, we only need to show that the expectation of $Z$ equals the latter quantity at first order. Notice that
    \begin{equation*}
    \begin{split}
    \mathbf{E}^{\rm DCM}[Z]&=\sum_{x,y\in\cV}\mathbf{E}^{\rm DCM}[\mathbf{P}^{\rm op}(c_x(t)=r\,,c_y(t)=r)]\\
&=\sum_{x,y\in\cV}\mathbf{E}^{\rm DCM}[\mathbf{P}_{x,y}^{\rm op}(\{x,y\}\subseteq\eta_t)]\,,
    \end{split}
    \end{equation*}
    where $\mathbf{P}_{x,y}^{\rm op}$ stands for the COBRAD process initialized with a particle at $x$ and a particle at $y$, and, we recall, the event $\left\{\{x,y\}\subseteq\eta_t\right\}$ means that both the labels survive up to time $t$. Arguing again as above we obtain that   \begin{equation}
\mathbf{E}^{\rm DCM}[Y^2]=(1+o(1))\mathbf{E}^{\rm DCM}[Y]^2\,,
\end{equation}
    and the desired result follows from \eqref{eq:30}.
\end{proof}

\bibliographystyle{alpha}
\bibliography{references}

\end{document}